\documentclass[12pt,preprint]{aastex}
\shorttitle{Gravitational Radiation from White Dwarfs}
\shortauthors{Benacquista, Sedrakian, {\it et al.}}
\begin{document}

\title{Gravitational Radiation from Pulsating White Dwarfs}
\author{M. Benacquista}
\affil{Montana State University-Billings, Billings, MT 59101}
\email{benacquista@msubillings.edu}

\and

\author{D.M. Sedrakian, M.V. Hairapetyan, K.M. Shahabasyan, and A.A. Sadoyan}
\affil{Yerevan State University, 375025 Yerevan, Armenia}
\email{dsedrak@www.physdep.r.am}
\email{mhayr@www.physdep.r.am}
\email{kshabas@www.physdep.r.am}
\email{asadoyan@www.physdep.r.am}

\begin{abstract}
Rotating white dwarfs undergoing quasi-radial oscillations can emit gravitational radiation in a frequency range from $0.1 - 0.3~{\rm Hz}$. Assuming that the energy source for the gravitational radiation comes from the oblateness of the white dwarf induced by the rotation, the strain amplitude is found to be $\sim 10^{-27}$ for a white dwarf at $\sim 50~{\rm pc}$. The galactic population of these sources is estimated to be $\sim 10^7$, and may produce a confusion limited foreground for proposed advanced detectors in the frequency band between space-based and ground-based interferometers. Nearby oscillating white dwarfs may provide a clear enough signal to investigate white dwarf interiors through gravitational wave asteroseismology.
\end{abstract}


\section{Introduction}
There are a number of gravitational radiation detectors, planned, under construction, and operational covering a wide frequency spectrum from $\sim 10^{-9}~{\rm Hz}$ all the way up to $\sim 10^{4}~{\rm Hz}$. Expected sources of gravitational radiation include numerous astrophysical sources such as compact binaries, supermassive black holes, and binary coalescence. In addition, there is an expected cosmological background of gravitational radiation arising from the very earliest times of the universe. The coverage of the spectrum is not complete and the gap between space-based interferometers (such as LISA) and ground-based interferometers (such as LIGO and VIRGO) has been proposed as a possible ``clean window'', devoid of continuous foreground sources, through which the cosmological background of gravitational radiation could be seen~\citep{seto01}. Most binary white dwarf systems will coalesce before their gravitational wave frequency rises above $0.1~{\rm Hz}$, while more massive binaries such as double neutron star or black hole binaries will be sweeping through this band on the way to their eventual coalescence in the ground-based frequency band. However, rotating white dwarfs undergoing quasi-radial oscillations will emit gravitational radiation in this frequency band. These sources will be essentially monochromatic and long-lived. Given the number of white dwarfs in the galaxy, it is quite possible that this population will produce a confusion-limited foreground of sources in this frequency band that will mask the cosmological background. We propose a possible energy source for these oscillations and attempt to estimate the average signal strength.

\section{Quasi-radial Oscillations}
Quasi-radial oscillations of rotating white dwarfs were investigated in the early 1970's~\citep{papoyan72, haroutyunyan72} where the frequency spectrum of the fundamental oscillation mode for maximally rotating white dwarfs was determined. These stars are oblate due to their rotation and consequently they have a non-zero quadrupole moment. The oscillations add a time dependence to the quadrupole moment~\citep{vartanian77}. The oscillation is described by assigning each mass element a time dependent coordinate given by $x_\alpha = x_\alpha^0(1+\eta \sin{\omega t})$ where $\eta \ll 1$ and a constant. Thus, the reduced quadrupole moment is given by:
\begin{eqnarray}
Q_{\alpha \beta} & = & \int{\rho\left( x_\alpha x_\beta - \onethird \delta_{\alpha \beta} x^2\right) d^3x} \nonumber \\
 & \simeq & Q_{\alpha \beta}^0\left(1+2\eta \sin{\omega t}\right)
\end{eqnarray}
where $Q_{\alpha \beta}^0$ are the components of the quadrupole moment of the rotating oblate white dwarf in equilibrium and we have neglected terms of order $\eta^2$. Taking the axis of rotation to lie along the $z$-axis, the non-zero components of the quadrupole moment obey:
\begin{equation}
Q^0 = -Q_{zz}^0 = 2Q_{xx}^0 = 2Q_{yy}^0.
\end{equation}

The power emitted in gravitational radiation is given by:
\begin{equation}
J = \frac{G}{5c^5}\left|\frac{d^3}{dt^3}Q_{\alpha \beta}\right|^2,
\end{equation}
and consequently one obtains:
\begin{equation}
J = \frac{6G}{5c^5}\eta^2\omega^6\left|Q^0\right|^2 \cos^2{\omega t^\prime} = J_0 \cos^2{\omega t^\prime}
\label{power}
\end{equation}
where the retarded time is $t^\prime = t - r/c$ for a source at distance $r$.

To determine the wave form and the angular distribution of the radiation, we rotate to coordinates in which the wave vector lies along the $z$-axis and use the transverse-traceless gauge. Consequently,
\begin{eqnarray}
h_+ & = & \onehalf\left(h_{xx} - h_{yy}\right) \nonumber \\
& = & \frac{3 G Q^0 \eta \omega^2}{c^4 r} \sin^2{\theta}\sin{\omega t^\prime} 
\label{hplus} \\
h_{\times} & = & h_{xy} = 0,
\label{htimes}
\end{eqnarray}
where $\theta$ is the angle between the wave vector and the axis of rotation of the white dwarf. We can express the strain amplitude in terms of the power by combining Eq.~\ref{power} with Eq.~\ref{hplus} to obtain:
\begin{equation}
h_+ = \sqrt{\frac{15GJ_0}{2c^3}} \frac{1}{r\omega} \sin^2{\theta} \sin{\omega t^\prime}.
\label{strain}
\end{equation}
If an energy source can be found to drive the pulsations, the rate at which power is put into the vibrations can be combined with the lifetime of the energy source to estimate the strain amplitude from an individual white dwarf and thus from the galactic population as a whole. We discuss a possible mechanism in the next section.

\section{Energy Source}
If there is no permanent source of energy to feed the gravitational radiation, the oscillation energy will quickly radiate away in about 1000 years~\citep{vartanian77}. Since the ultimate source of the gravitational radiation from the white dwarf is the oblateness arising from the rotation, we propose that the deformation energy of the white dwarf provides the energy to drive the oscillations. In this scenario, as the white dwarf spins down, it will transition from oblate to spherical. This transition will trigger starquakes which will feed the oscillations which drive the gravitational radiation. We assume that a part of the deformation energy will be converted into gravitational radiation while the remaining part will dissipate through thermal, electromagnetic, and other channels. We use the technique of~\citet{sahakian72} to calculate the deformation energy.

From numerical results, the dependence of the mass of the rotating and non-rotating configurations is found to be linear with the baryon number, so $M = kN$ and $M_0 = k_0N$, where the subscript ``0'' indicates the non-rotating configuration. The mass difference between the rotating and non-rotating configurations with the same baryon number is a result of the additional energy of rotation ($W_r(\Omega) = \onehalf I \Omega^2$) as well as the potential energy due to crustal deformation ($W_g(\Omega)$), thus:
\begin{equation}
W_g(\Omega) = \Delta M c^2 - W_r(\Omega),
\end{equation}
where $\Omega$ is the angular velocity of the white dwarf, $I$ is its moment of inertia, and the mass difference in grams is given by
\begin{equation}
\Delta M = (k-k_0)N = 8.96 \times 10^{-29} N.
\end{equation}
The appropriate parameters for maximally rotating white dwarfs were calculated in~\citet{haroutyunian71} and~\citet{sahakian72}. These parameters are presented in Table~\ref{wdparameters}. To obtain the deformation energy for rotation rates less than $\Omega_{\rm max}$, we use the fact that the results of~\citet{haroutyunian71} were obtained using a linear expansion in the small dimensionless parameter $\Omega^2/8\pi G \rho_c$, where $\rho_c$ is the non-rotating central density. Consequently, we can write:
\begin{equation}
W_g(\Omega) = \left(\frac{\Omega}{\Omega_{\rm max}}\right)^2W_g(\Omega_{\rm max}).
\end{equation}
It now remains to determine the time scale, $\tau$, for a spin-down mechanism so that we can relate the power in gravitational radiation to the decrease in deformation energy by:
\begin{equation}
J_0 = \beta \frac{W_g}{\tau},
\label{jocalc}
\end{equation}
where $\beta \ll 1$ is a ``branching ratio'' that quantifies the fraction of deformation energy which goes into gravitational radiation from the fundamental mode.

One possible mechanism for the spin-down of a rotating white dwarf is magneto-dipole radiation torque, which occurs if the magnetic field is oblique~\citep{schmidt01}. Observational data for 65 isolated white dwarfs indicates the magnetic field strength on the surface of these stars lies in the range $\sim 3\times 10^4$ to $\sim 10^9$ G \citep{wickramasinghe00}. If we define $\alpha$ to be the angle between the magnetic and rotation axes, the spin-down rate of the white dwarf is given by:
\begin{equation}
\dot{\Omega} = - \frac{2 \mu^2 \Omega^3}{3 I c^3} \sin^2{\alpha},
\label{omegadot}
\end{equation}
where $\mu = B R^3$ is the magnetic moment, $B$ is the magnetic field strength, and $R$ is the radius of the white dwarf. The characteristic time scale is then:
\begin{equation}
\tau = \frac{\Omega}{2\left|\dot{\Omega}\right|}.
\label{taucalc}
\end{equation}
We neglect the influence of gravitational radiation on the spin evolution because the ratio of the energy lost through gravitational radiation to that of the energy lost through the magnetic-dipole spin down mechanism is about $3.6 \beta$. With this information, we can now estimate the gravitational radiation luminosity and strain amplitude for several white dwarfs.

\section{Strain Amplitudes and Luminosities}
The rotation rates for white dwarfs are difficult to measure since these objects have little or no surface blemishes and gravitational broadening of their spectral lines overwhelms the expected rotational broadening~\citep{kawaler03}. Fortunately, some magnetic white dwarfs show time variability in their magnetic features which allows for their rotation rates to be inferred~\citep{wickramasinghe00,kawaler03}. Magnetic white dwarfs are thought to make up roughly 1\% of the white dwarf population and isolated magnetic white dwarfs tend to have an average mass ($\ga 0.95 M_{\sun}$) that is well above that of nonmagnetic white dwarfs ($\sim 0.57 M_{\sun}$). There is evidence that the population of magnetic white dwarfs is bimodal with high mass stars (near the Chandrasekhar limit) having rotational periods between $\sim 700$ s and several hours~\citep{wickramasinghe00}. This may be evidence that such objects are the result of double degenerate mergers~\citep{wickramasinghe00, dupuis02}, as has been proposed for EUVE J0317-855~\citep{ferrario97}.

The population of merged magnetic white dwarfs may be the most promising source of gravitational radiation from the mechanism of quasi-radial oscillations. To estimate the expected strain amplitude from both individual sources and the population as a whole, we identify eight candidates from the 65 known isolated magnetic white dwarfs from~\citet{wickramasinghe00} as those with rotational periods less than one day. The properties of the candidates are listed in Table~\ref{imwds}. We estimate the local space density of merged magnetic white dwarfs by taking the local space density of white dwarfs to be $0.003~{\rm pc}^{-3}$~\citep{liebert88}, the fraction of isolated magnetic white dwarfs to be $5.1 \%$~\citep{wickramasinghe00}, and the fraction of isolated magnetic white dwarfs that are remnants of mergers to be $8/65$. This results in a local space density of $\rho_{\rm s}=1.9 \times 10^{-5}~{\rm pc}^{-3}$. The expected distance to the nearest source is then found by:
\begin{equation}
r = 2 \left(\frac{3}{4\pi \rho_{\rm s}}\right)^{1/3} \simeq 46~{\rm pc}
\label{raverage}
\end{equation}
Assuming a galactic distribution of white dwarfs to follow the disk population, we assign a density distribution of:
\begin{equation}
\rho = \rho_0 e^{-r/R} e^{-z/h}
\end{equation}
in galacto-centric cylindrical coordinates, with $R = 2.5~{\rm kpc}$ and $h = 200~{\rm pc}$. Taking the solar location as $r_s = 8.5 {\rm kpc}$ and $z_s = 0$, we obtain $\rho_0 = 5.5 \times 10^{-4}~{\rm pc}^-3$ and a total number of $N = 8.6\times10^6$ in the galaxy.

Expected strain amplitudes, $\tau$, and $\eta$ for the eight isolated magnetic white dwarfs are given in Table~\ref{strainamp}. These are calculated by first determining $\tau$ from Eqs.~\ref{omegadot} and~\ref{taucalc}, where we have chosen $\sin^2{\alpha} = 1/2$. This is a reasonable assumption since a distribution in $\alpha$ that is either uniform or spiked at $\alpha = \pi/4$ are supported by observation~\citep{schmidt91}. Choosing a value for $\beta$ with which to calculate the luminosity is somewhat problematic due to the dearth of observational data for pulsations in this frequency range. Most observations of white dwarf pulsations are in the mHz range, and consequently any pulsations in the dHz range will be averaged out in an observation. We choose an upper bound to $\beta$ by requiring that the largest Doppler broadening of spectral lines due to pulsations be less than thermal Doppler broadening. The most stringent constraint comes from PG 1031+234 and yields $\beta = 10^{-4}$. We note that the resulting pulsational Doppler broadening in the remaining seven white dwarfs is at least an order of magnitude below the thermal Doppler broadening. Thus, the luminosity is calculated from Eq.~\ref{jocalc} and used with Eqs.~\ref{raverage} and~\ref{strain} to determine the strain amplitude where we have averaged over all orientations and taken $\beta = 10^{-4}$. We have used the average mass of $0.95~M_{\sun}$ whenever the mass was undetermined and the average distance of $46~{\rm pc}$ is the distance is unknown. The expected energy flux on earth (in ${\rm erg}~{\rm s}^{-1}~{\rm cm}^{-2}$) for a population made entirely of each type of white dwarf is also shown in Table~\ref{strainamp}. The flux, $F$, is calculated using
\begin{equation}
F = \frac{4\pi \rho_{\rm s} h f(h) J_0}{4 \pi \left(3 \times 10^{18}\right)^2}
\end{equation}
where $h = 200~{\rm pc}$ and $f(h) = 6.15$ is calculated using Appendix A of~\citet{hils90}. Finally, we note that a simple average of the strain amplitudes in Table~\ref{strainamp} gives $h_+ = 5.9 \times 10^{-28}$ and an average flux of $F = 9.21 \times 10^{-16}~{\rm erg}~{\rm s}^{-1}~{\rm cm}^{-2}$. The flux is spread out over a frequency band of $\nu_1=0.12$ to $\nu_2=0.32~{\rm Hz}$, and we can estimate an average strain amplitude for the galactic population of pulsating white dwarfs by using the angle and polarization averaged expression of~\citet{douglas79} and averaging over the frequency range $\Delta \nu = \nu_2 - \nu_1$ to obtain:
\begin{equation}
h_{+ \rm ave} = \frac{\ln{\nu_2/\nu_1}}{\Delta \nu} \sqrt{\frac{4 G F}{\pi c^3}},
\end{equation}
which gives $h_{+ \rm ave} = 8.35 \times 10^{-27}$.

\section{Conclusions}

We have shown that the galactic population of white dwarfs can produce a background of gravitational radiation in the frequency range of $0.12-0.32~{\rm Hz}$ through quasi-radial pulsations. The source of energy to drive these pulsations is found in the deformation energy of the white dwarf due to its rotation. This energy can be extracted from the white dwarf as it spins down. We have proposed that a population of isolated magnetic white dwarfs which are the remnants of merged double degenerate binaries can be sources of this gravitational radiation. Estimates of the signal strength over the frequency band of interest indicate that this population may be comparable in strength to the level of the stochastic cosmological background of gravitational radiation predicted by standard inflationary models. In addition, some white dwarfs may be near enough that their signals will stand above the background and their individual parameters can be measured. If the frequency and amplitude of the gravitational radiation from one of these nearby single white dwarfs can be measured, then it will provide an opportunity to test models of white dwarf interiors and give a unique opportunity to check the main parameters for white dwarfs and to understand their dissipation mechanisms.

\acknowledgments
This work is supported by CRDF award AP2-3207 and 12006/NFSAT PH067-02.  MJB is also supported by NASA Cooperative Agreement NCC5-579.

\begin{deluxetable}{rrrrrrrrrrr}
\tablecaption{Structural Parameters of Maximally Rotating White Dwarfs \label{wdparameters}}
\tablehead{\colhead{$\rho_{c(6)}$} & \colhead{$N_{(57)}$} & \colhead{$M/M_{\sun}$} & \colhead{$\Omega_{\rm max}$} & 
\colhead{$I_{(48)}$} & \colhead{$Q^0_{(48)}$} & \colhead{$R_{(8)}$} & \colhead{$\Delta M c^2_{(49)}$} & \colhead{$W_{r(49)}$} & \colhead{$W_{g(49)}$} & \colhead{$\omega$}}
\startdata
2.403 & 0.4997 & 0.5945 & 0.196 & 128.0\phn & 20.48\phn & 10.93\phn & 4.799 & 0.246 & 4.55 & 0.758\\
19.38\phn & 0.8398 & 0.9993 & 0.476 & 88.6\phn & 14.27\phn & 7.342  & 8.066 & 1.00\phn & 7.06 & 0.794\\
157.7\phn\phn & 1.0695 & 1.2731 & 1.063 & 39.5\phn & 4.766 & 4.625  & 10.27\phn & 2.23\phn & 8.05 & 1.513\\
866.1\phn\phn & 1.1340 & 1.3502 & 2.042 & 15.9\phn & 1.554 & 3.044 & 10.90\phn & 3.32\phn & 7.58 & 1.985\\
2586\phd\phn\phn\phn & 1.1261 & 1.3412 & 3.105 & 8.17 & 0.673 & 2.287 & 10.82\phn & 3.94\phn & 6.89 & 0.967
\enddata
\tablecomments{The values of $\rho_{c(6)}$ are in units of $10^6 {\rm g}/{\rm cm}^3$; $N_{(57)}$ is in units of $10^{57}$ particles; $I_{(48)}$ and $Q^0_{(48)}$ are in units of $10^{48} {\rm g}\cdot{\rm cm}^2$; $R_{e(8)}$ is the equatorial radius in $10^8~{\rm cm}$; $\Delta M c^2_{(49)}$, $W_{r(49)}$, and $W_{g(49)}$ are in units of $10^{49}$ erg.}
\end{deluxetable}

\begin{deluxetable}{lrrcr}
\tablecaption{Isolated Magnetic White Dwarfs \label{imwds}}
\tablehead{\colhead{Name} & \colhead{$r$ (pc)} & \colhead{$B$ (MG)} & \colhead{$P_{\rm rot}$} & \colhead{$M/M_{\sun}$}}
\startdata
PG 1031+234 & 142\tablenotemark{a} & 500\tablenotemark{b} & 3.4 hr & 0.93\tablenotemark{c} \\
EUVE J0317-855 & 35\tablenotemark{a} & 450 & 725 s & 1.35\\
PG 1015+015 & 66\tablenotemark{a} & 90 & 99 min & 1.15\tablenotemark{c}\\
Feige 7 & 49\tablenotemark{a} & 35 & 2.2 hr & 0.6\\
G99-47 & 8\tablenotemark{c} & 25 & 1 hr? & 0.71\tablenotemark{c}\\
KPD 0253+5052 & 81\tablenotemark{d} & 17 & 3.79 hr & \nodata\\
PG 1312+098 & \nodata & 10 & 5.43 hr & \nodata\\
G217-037 & 11\tablenotemark{c} & $\lesssim 0.2$ & 2-20? hr\tablenotemark{d} & 0.89
\enddata
\tablenotetext{a}{~\citet{heyl00}.}
\tablenotetext{b}{$B$ ranges from 500 to 1000.}
\tablenotetext{c}{~\citet{liebert03}.}
\tablenotetext{d}{~\citet{downes86}.}
\tablenotetext{e}{$P_{\rm rot} = 2$ hr was used for the calculation.}
\end{deluxetable}

\begin{deluxetable}{lrrrr}
\tablecaption{Strain Amplitudes and Fluxes for Isolated White Dwarfs with $\beta = 10^{-4}$. \label{strainamp}}
\tablehead{\colhead{Name} & \colhead{$h_+$} & \colhead{$F$} & \colhead{$\tau$ (Gyr)} & \colhead{$\eta$}}
\startdata
PG 1031+234 & $6.0 \times 10^{-29}$ & $6.1 \times 10^{-17}$ & 11 & $1.0 \times 10^{-2}$ \\
EUVE J0317-855 & $1.0 \times 10^{-27}$ & $6.7 \times 10^{-15}$ & 1.7 & $4.0 \times 10^{-3}$ \\
PG 1015+015 & $9.3 \times 10^{-30}$ & $1.1 \times 10^{-18}$ & 571 & $7.1 \times 10^{-4}$ \\
Feige 7 & $1.6 \times 10^{-28}$ & $4.9 \times 10^{-17}$ & 125 & $5.2 \times 10^{-4}$ \\
G99-47 & $3.5 \times 10^{-27}$ & $5.9 \times 10^{-16}$ & 50.6 & $3.7 \times 10^{-4}$ \\
KPD 0253+5052 & $2.9 \times 10^{-30}$ & $4.6 \times 10^{-20}$ & 11,845 & $3.5 \times 10^{-4}$ \\
PG 1312+098 & $1.5 \times 10^{-30}$ & $3.8 \times 10^{-21}$ & 70,266 & $2.0 \times 10^{-4}$ \\
G217-037 & $9.0 \times 10^{-31}$ & $8.2 \times 10^{-23}$ & $2.4 \times 10^{7}$ & $4.1 \times 10^{-6}$
\enddata
\end{deluxetable}

\end{document}